\documentclass[12pt]{iopart}
\usepackage{graphicx}

\begin{document}
\title{Thermodynamics of histories for the one-dimensional contact process}

\author{Jef Hooyberghs$^{1,2}$ and Carlo Vanderzande$^{2,3}$}

\address{$^1$VITO, Boeretang 200, 2400 Mol, Belgium}
\address{$^2$Departement WNI, Hasselt University, 3590 Diepenbeek, Belgium}
\address{$^3$Instituut Theoretische Fysica, Katholieke Universiteit Leuven, 3001 Heverlee, Belgium}
\ead{carlo.vanderzande@uhasselt.be}
\begin{abstract}
The dynamical activity $K(t)$ of a stochastic process is the number of times it changes configuration up to time $t$. It was recently argued that (spin) glasses are at a first order dynamical transition where histories of low and high activity coexist.
We study this transition in the one-dimensional contact process by weighting its histories by $\exp(sK(t))$. 
We determine the phase diagram and the critical exponents of this model using a recently developed approach to the thermodynamics of histories that is based on the density matrix renormalisation group. We find that for every value of the infection rate, there is a phase transition at a critical value of $s$. 
Near the absorbing state phase transition of the contact process, the generating function of the activity shows a scaling behavior similar to that of the free energy in an equilibrium system near criticality. \\


\end{abstract}

\maketitle
\section{Introduction}
While phase transitions in equilibrium can only occur in two or higher dimensions, it is by now well established that they can be present in non equilibrium steady states even in one dimension \cite{Marro99}. Examples can be found in driven lattice gases and in models showing a transition out of an absorbing state. In the latter case, the system changes from an absorbing state to an active one as a parameter is varied. In the absorbing phase, the system is frozen in a particular configuration which it never leaves: its dynamical activity $K(t)$, which equals the number of configuration changes up to some time $t$, becomes therefore zero (after relaxation to the absorbing state). In the active phase, the system changes state regularly and $K(t)$ becomes nonzero.
In this paper, we investigate the statistical properties of the activity in the contact process, a well known process showing an absorbing state phase transition. 
We do this within the formalism of thermodynamics of histories which was recently adapted to Markov processes \cite{Lecomte07}. 

Our study was motivated by recent work on glasses. Indeed, it has been argued that  they show 'dynamical coexistence' between histories of high and of low activity \cite{Merolle05,Jack06} and that they therefore are at a dynamical first-order transition. This picture was confirmed in a recent study \cite{Garrahan07,Garrahan09} of one dimensional kinetically constrained models of glasses \cite{Ritort03}. These models have a dynamics with detailed balance and therefore show no equilibrium transition in one dimension. Yet, when their histories are weighted with $e^{sK(t)}$, clear evidence for a first-order dynamical transition at $s=0$ was found. More recently, a similar transition was found in more realistic, three-dimensional, atomistic models of glass formers \cite{Hedges09} where it occurs at a small but nonzero value of $s$. In a spin-glass model, the transition occurs at $s=0$ below some temperature but at a non-zero $s$ above that temperature \cite{Jack09}. Here we present evidence that in this respect, the one dimensional contact process behaves like a higher dimensional glass or spin glass. We find evidence for a dynamical transition at $s > 0$ when the contact process is in the absorbing state, and at $s=0$ in the active phase. Near the absorbing/active transition the cumulant generating function of the activity show an interesting scaling behaviour. We obtain our results using a density matrix renormalisation group (DMRG) approach to the thermodynamics of histories that we introduced recently \cite{Gorissen09}.

While the contact process is not a model for a glass, an extended version of the Fredrickson-Andersen model of glasses, shows a transition in the same universality class as the contact process \cite{Jack06} (the directed percolation class \cite{Grassberger79}). Our results on the scaling of the activity may therefore also hold in glass models.

Besides its relevance for glasses, the dynamical activity has in recent years gained importance in non equilibrium statistical mechanics \cite{Lecomte07,Maes06} since it appears in several contexts, such as in a generalization of the fluctuation-dissipation theorem to non equilibrium steady states \cite{Baiesi09}.
The activity is symmetric under time reversal and therefore plays a complementary role to entropy production which is anti-symmetric \cite{Maes06}.

This paper is organised as follows. In section 2 we introduce the contact process and show how the statistical properties of its activity can be determined from the largest eigenvalue of a generalised generator. 
We also show that a dynamical transition, if present, should occur at $s\geq 0$. In section 3, we briefly explain the use of the DMRG for these kind of problems. Our results on the phase diagram are discussed in section 4. In section 5 we present results on critical exponents and scaling. Our conclusions are presented in section 6. Some details on the DMRG technique used are given in the appendix.

\section{The activity of the contact process}
In the contact process \cite{Harris74}, each site of a lattice $\Lambda$ can be occupied by at most one particle. Introduced as a model in epidemiology, a particle is associated with a sick person, whereas a vacancy corresponds to a healthly individual. It is thus convenient to use a spin language where an empty (occupied) site corresponds to a spin up (down). A microstate $C$ of the system is then given by the set $\{\sigma_i^z, i \in \Lambda\}$. In this paper, we restrict ourselves to one-dimensional lattices of $L$ sites with open boundaries.

The dynamics of the contact process is a continuous time Markov process for which 
the probability $P(C,t)$ that the system is in a microstate $C$ at time $t$ evolves according to the master equation
\begin{eqnarray}
\partial_t P(C,t) = \sum_{C' \neq C} \left[w(C' \to C) P(C',t) - w(C \to C') P(C,t)\right]
\label{meq}
\end{eqnarray}
Here $w(C \to C')$ is the transition rate to go from  $C$ to $C'$. For the contact process two types of transition are allowed: an occupied site can become empty with rate 1, while an empty site becomes occupied with rate $\zeta \lambda/2$. Here $\zeta$ is the number of occupied neighbours and $\lambda$, which is the only parameter in the model, is  the infection rate (within an epidemiological context).
For further reference, we introduce the inverse lifetime $r(C)$ of the state $C$ which is given by $r(C)=\sum_{C' \neq C} w(C \to C')$.  

The contact process is a standard model for phase transitions out of an absorbing state \cite{Marro99}.
When $\lambda<\lambda_c$, the contact process reaches an absorbing state in which all sites are empty. For $\lambda>\lambda_c$ and in an infinite system, the model goes to a stationary state with a finite density $\rho$ of particles.  It is known from extensive numerical investigations that this phase transition belongs to the universality class of directed percolation \cite{Grassberger79}. The scaling properties of various quantities near $\lambda_c$ are well characterised and precise numerical values for the critical exponents are known, especially in one dimension \cite{Marro99}. In finite systems, the stationary state will always be the absorbing one. It is therefore common to add as an extra process the creation of particles at the boundary sites with a rate $\omega$. This addition is not believed to influence the critical behavior of the model. In our calculations, we choose $\omega=1$.

In this paper, we are interested in {\it the dynamical activity} $K(t)$ of the contact process. It gives the number of times that the system has changed configuration in a particular realisation (history)  of the process, and this up to time $t$. Mathematically, the statistical properties of the activity can be obtained from its (cumulant) generating function
\begin{eqnarray}
\pi(s,\lambda,L)\equiv \lim_{t \to \infty} \frac{1}{t} \ln\langle e^{sK(t)}\rangle
\label{2}
\end{eqnarray}
where the average is taken over the histories of the stochastic process. We have explicitly added the $L$-dependence because we will later on formulate a finite size scaling theory for $\pi$. Equation (\ref{2}) shows a large mathematical similarity with the definition of the free energy in equilibrium in which the inverse temperature $\beta=1/k_BT$ is replaced by $(-s)$ and the sum over configurations is replaced by a sum over histories. 
Hence, the name thermodynamics of histories. As we will show below, $\ln \langle e^{sK_L(t)}\rangle$ is extensive in time, in contrast with the free energy which is extensive in volume.

The probability distribution of $K(t)$ at large times can be determined from $\pi(s,\lambda,L)$ through a Legendre transformation \cite{Derrida07}, while the average activity $K(\lambda,L)$, its variance $\Delta_K(\lambda,L)$, and higher cumulants can be found as derivatives of $\pi(s,\lambda,L)$:
\begin{eqnarray}
K(\lambda,L)=\lim_{t \to \infty} \frac{1}{t} \langle K(t)\rangle = \frac{\partial \pi}{\partial s}(0,\lambda,L) \label{3} \\
\Delta_K(\lambda,L) = \lim_{t \to \infty} \frac{1}{t} \left[\langle K^2(t)\rangle-\langle K(t)\rangle^2\right]=\frac{\partial^2 \pi}{\partial s^2}(0,\lambda,L) \label{4}
\end{eqnarray}
The variable $s$ can be seen as a kind of fugacity that weights histories of the contact process according to their activity. For $s \ll 0$, non-active histories are favoured, whereas for $s\gg 0$, very active histories determine $\pi$. In this paper, we are interested in the behaviour of the histories of the contact process as a function of $\lambda$ and $s$. 
While the parameter $s$ has no clear "physical" interpretation, we will show that the one-dimensional contact process has a phase transition as a function of $s$, for all $\lambda$-values. This transition is dynamic in nature, and separates regions in parameter space where there is no activity from regions with activity. This allows us to see the absorbing state phase transition in the contact process in a broader context. Moreover, as argued in the introduction the behaviour found here in one dimension could be a non equilibrium version of a phenomenon that occurs in (spin) glasses in two or more dimensions.

For clarity, we now repeat a standard argument \cite{Lecomte07} that shows that $\pi(s,\lambda,L)$ equals the largest eigenvalue of a matrix $H(s)$. We therefore introduce firstly the probability $P(C,K,t)$ that the system is in configuration $C$ and has activity $K$ at time $t$. Using (\ref{meq}), we immediately find that
\begin{eqnarray}
\partial_t P(C,K,t) = \sum_{C'\neq C} \left[w(C'\to C) P(C',K-1,t)-w(C\to C') P(C,K,t)\right] \nonumber \\
\label{4}
\end{eqnarray}
Consequently, the discrete Laplace transform $\hat{P}(C,s,t)=\sum_{K=0}^{\infty} e^{sK} P(C,K,t)$ evolves according to
\begin{eqnarray}
\partial_t \hat{P}(C,s,t) = \sum_{C'\neq C} \left[ w(C'\to C) e^{s} \hat{P}(C',s,t)-w(C\to C') \hat{P}(C,s,t)\right]
\label{5}
\end{eqnarray}
To continue, it is convenient to introduce a matrix notation as common in the so called "quantum" approach to stochastic particle systems \cite{Schutz00}. We therefore introduce a set of basis vectors $| C\rangle$ each corresponding to a microstate $C$ and a vector $|\hat{P}(s,t)\rangle$ with components $\hat{P}(C,s,t)=\langle C|\hat{P}(s,t)\rangle$. Using this notation, the set of equations (\ref{5}) can be rewritten as
\begin{eqnarray}
\partial_t |\hat{P}\rangle = H(s) |\hat{P}\rangle
\label{6}
\end{eqnarray}
The diagonal elements of the matrix $H(s)$ are equal to minus  the inverse lifetimes of the states, while the off-diagonal elements are given by the transition rates multiplied by $e^s$. For $s=0$, (\ref{6}) reduces to the master equation (\ref{meq}) and $H(0)$ corresponds to the generator of the stochastic process. We will therefore refer to $H(s)$ as the generalised generator. 

Using the "quantum" notation of \cite{Schutz00} one can easily show that for the contact process
\begin{eqnarray}
H(s) &=& \sum_{i=1}^L \left[ (e^s s_i^+ -n_i) + \frac{\lambda}{2}(n_{i-1}+n_{i+1})(e^s s_i^- - v_i)\right] \nonumber \\ &+& \omega (s_1^- + s_L^--v_1-v_L)
\label{6a}
\end{eqnarray}
$(n_0=n_{L+1}=0)$. Here $n_i, v_i, s_i^+$ and $s_i^-$ are standard particle number, vacancy number, particle annihilation and creation operators at site $i$
\begin{eqnarray}
n=\left( \begin{array}{cc} 0\ \ 0 \\ 0\ \ 1 \end{array}\right),\ \
v=\left( \begin{array}{cc} 1\ \ 0 \\ 0\ \ 0 \end{array}\right),\ \
s^+=\left( \begin{array}{cc} 0\ \ 1 \\ 0\ \ 0 \end{array}\right),\ \
s^-=\left( \begin{array}{cc} 0\ \ 0 \\ 1\ \ 0 \end{array}\right)
\label{6b}
\end{eqnarray}

The formal solution to (\ref{6}) is
\begin{eqnarray}
|\hat{P}(s,t) \rangle = e^{H(s) t} |\hat{P}(s,0)\rangle
\label{7}
\end{eqnarray}
Therefore we have
\begin{eqnarray}
\langle e^{sK(t)}\rangle &=& \sum_C \sum_K e^{sK} P(C,K,t) \nonumber \\
&=& \sum_C \hat{P}(C,s,t) \nonumber \\
&=& \sum_C \langle C|e^{H(s) t} |\hat{P}(s,0)\rangle 
\label{7}
\end{eqnarray}
Using the spectral theorem the sum in (\ref{7}) can be written as a sum over the eigenvalues of $H(s)$. In the long time limit in which we are interested, this sum is dominated by the largest eigenvalue $\mu_0(s,\lambda,L)$ of the generalised generator, so that for large $t$ we write (\ref{7}) as
\begin{eqnarray}
\langle e^{sK(t)}\rangle =\left(\sum_C \langle C|v_0\rangle\langle v_0|\hat{P}(s,0)\rangle \right) e^{\mu_0 (s,\lambda,L) t}\left[1 + \cdots\right]
\label{8}
\end{eqnarray}
Here $|v_0\rangle$ is the eigenvector associated with $\mu_0$. Taking $t \to \infty$ the terms coming from smaller eigenvalues can be neglected and using (\ref{2}) we finally obtain
\begin{eqnarray}
\pi(s,\lambda,L) =\mu_0(s,\lambda,L)
\label{9}
\end{eqnarray}
The cumulant generating function of the activity therefore equals the largest eigenvalue of $H(s)$. In this way, we have also shown that $\ln \langle e^{sK(t)}\rangle$ is extensive in time. Hence, the definition (\ref{2}) makes sense.

Some insight on the behaviour of $\pi$ for negative $s$ can be obtained from simple arguments. First we observe that $\pi$ is non decreasing and by definition is zero at $s=0$. Hence we obtain the bounds
\begin{eqnarray}
\pi(s\to-\infty,\lambda,L)\leq \pi(s,\lambda,L)\leq 0 \hspace{2cm}Ês\leq 0
\label{91}
\end{eqnarray} 
For $s \to -\infty$, only histories with no activity contribute to $\pi$. This implies that the system at time $t$ still is in the microstate $C$ in which it was initially (at time $t=0$). The probability for this decays exponentially with waiting time $1/r(C)$ so that
\begin{eqnarray}
\lim_{s \to -\infty} \langle e^{sK}\rangle = \sum_C \exp{\left[-r(C) t\right]}
\end{eqnarray}
In order to determine $\pi$ we have to take the long time limit of this expression, which will be dominated by $\min_C r(C)$. Provided limits can be interchanged, we obtain
\begin{eqnarray}
\lim_{s \to -\infty} \pi(s,\lambda,L) = - \min_C r(C)
\label{92}
\end{eqnarray}
This result is quite general and holds also for other models. In the particular case of the contact process one has for the completely empty configuration, $r(C)=2\omega=2$. For configurations with one particle $r(C)=3+\lambda/2$ if the particle is in the bulk and $r(C)=2+\lambda/2$ if it is on the boundary. Continuing in this way, one can easily see that $\min_C r(C)=2$. Putting everything together we find
\begin{eqnarray}
-2 \leq \pi(s,\lambda,L)\leq 0 \hspace{2cm}Ês\leq 0
\label{93}
\end{eqnarray}
This also implies that the intensive quantities  $\pi(s,\lambda,L)/L$ and $K(\lambda,L)$ are constant and equal to zero for any negative $s$. Hence, if there is some non analytical behaviour in these quantities, it should occur at $s_c\geq 0$.

\section{DMRG approach to activity fluctuations}
To calculate the cumulant generating function of the activity of the contact process we have to calculate the largest eigenvalue of the generalised generator (\ref{6a}). If we interpret this generator as a 'Hamiltonian' \cite{Schutz00}, calculating its largest eigenvalue is mathematically similar to determining the ground state energy of a quantum spin chain. The main difference with a standard quantum problem lies in the fact that in the present case the 'Hamiltonian' is non-Hermitian.

One of the most succesfull numerical procedures to calculate ground state properties of quantum chains is the density matrix renormalisation group (DMRG) introduced by S White \cite{White92,White93} (for a review see \cite{Schollwock05}). 
Later, the technique was adapted for generators of stochastic processes \cite{Kaulke98,Carlon99}, and recently we applied it for the first time to the generalised generators associated with activity fluctuations in the contact process and with current fluctuations in driven lattice gases \cite{Gorissen09}.

The application of the DMRG to $H(s)$ is not fundamentally different from the standard approach used for quantum systems. For stochastic systems two modifications are necessary. The first one is straightforward: one needs  precise diagonalisation algorithms capable of handling non-Hermitian matrices. These exist (e.g. the Arnoldi algoritm \cite{Golub96}), but are much more time consuming than their Hermitian counterparts. They are also numerically less stable. This limits the sizes we can reach to $L\approx 60$. 

The second problem concerns the fact that left and right eigenvectors of $H(s)$ are not related by transposition. In Appendix A, we briefly discuss the procedure that we followed to solve this problem.

Once the largest eigenvalue has been calculated with sufficient numerical accuracy, the average activity and its variance are determined using numerical differentiation. Because of numerical errors, it is not possible to obtain cumulants beyond the third with high enough precision.
\section{The phase diagram}
In this section, we present evidence that there exist a phase transition between dynamically active and inactive phases for every $\lambda$.

Firstly, we show in Figure 1 (left side) a typical result for $\pi(s,\lambda,L)$. In this case $\lambda=4.3$, which is in the active phase ($\lambda_c\simeq3.2978$ in one dimension). As can be seen, $\pi$ tends to $-2$ for decreasing $s$-values, as predicted in section 2. There is only a weak $L$-dependence, which is most pronounced near $s=0$ as can be seen on the right side of Figure 1. 

From results like these we can calculate the $s$-weighted average activity and its variance as a function of $s$ for various $L$-values. In Figure 2, we show our results, again for an infection rate above the critical one. As can be seen, a peak occurs in the variance. The height of the peak grows with $L$, which is a strong indication of the occurence of a phase transition. The average activity has a steep increase at this point. 

\begin{figure}[t]
\includegraphics[width=8.0cm]{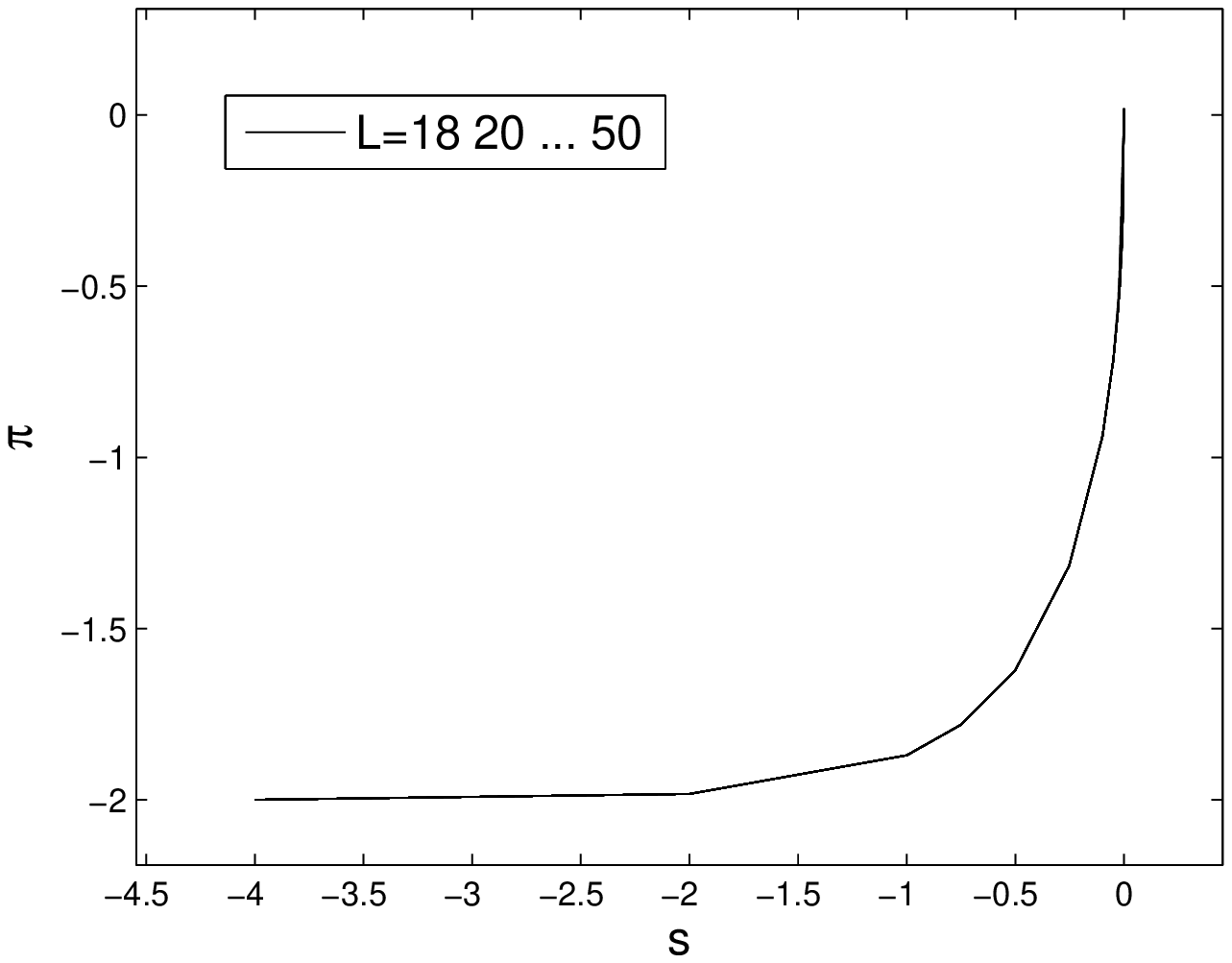}\includegraphics[width=8.0cm]{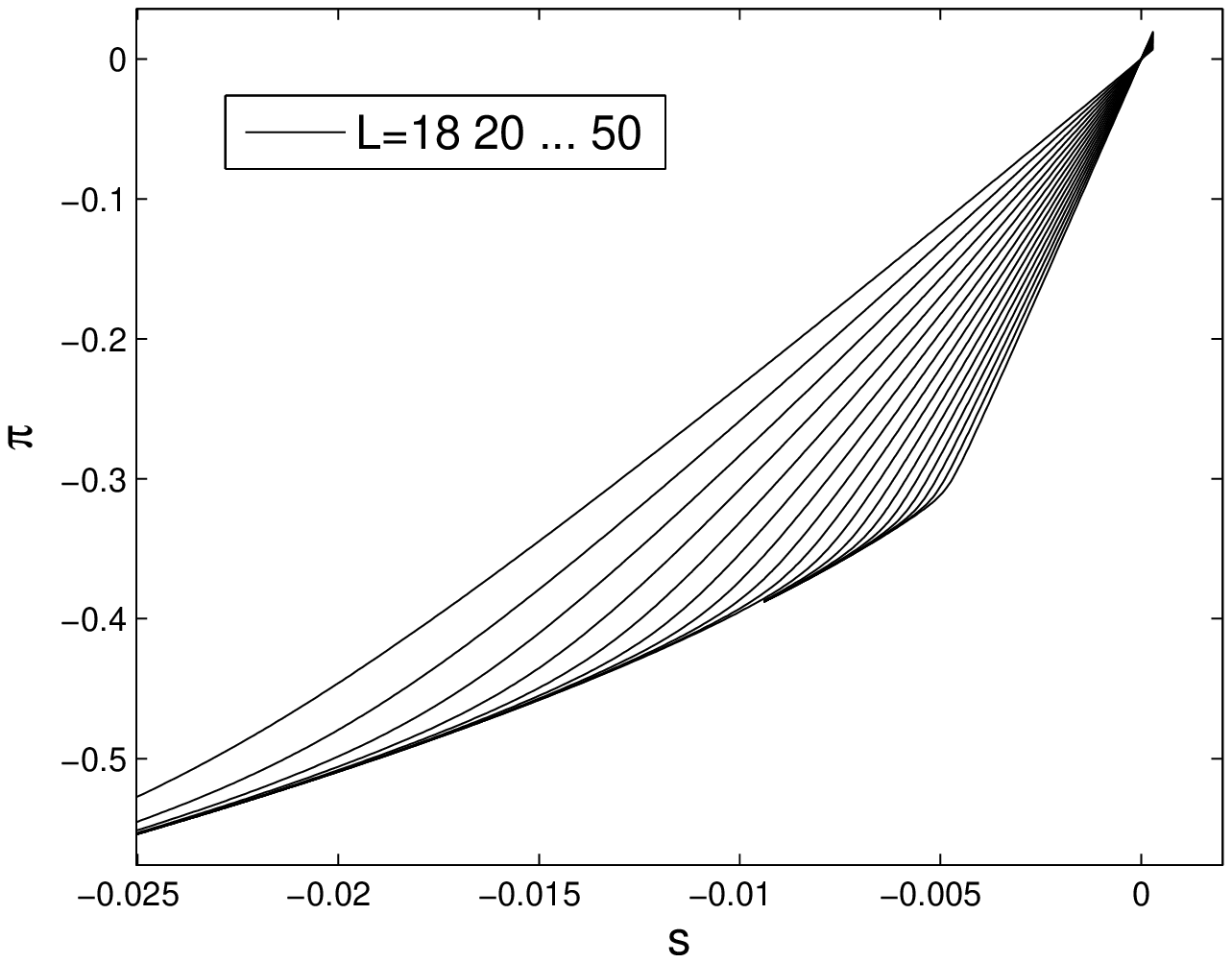}
\caption{\label{fig1} Cumulant generating function for $\lambda=4.3,s\leq 0$ for $L=18,20,...,50$. On the scale of the left figure, one cannot see any $L$-dependence. On the right, the region close to $s=0$ is enlarged. System size increases from top to bottom.}
\end{figure}

\begin{figure}[t]
\includegraphics[width=8.0cm]{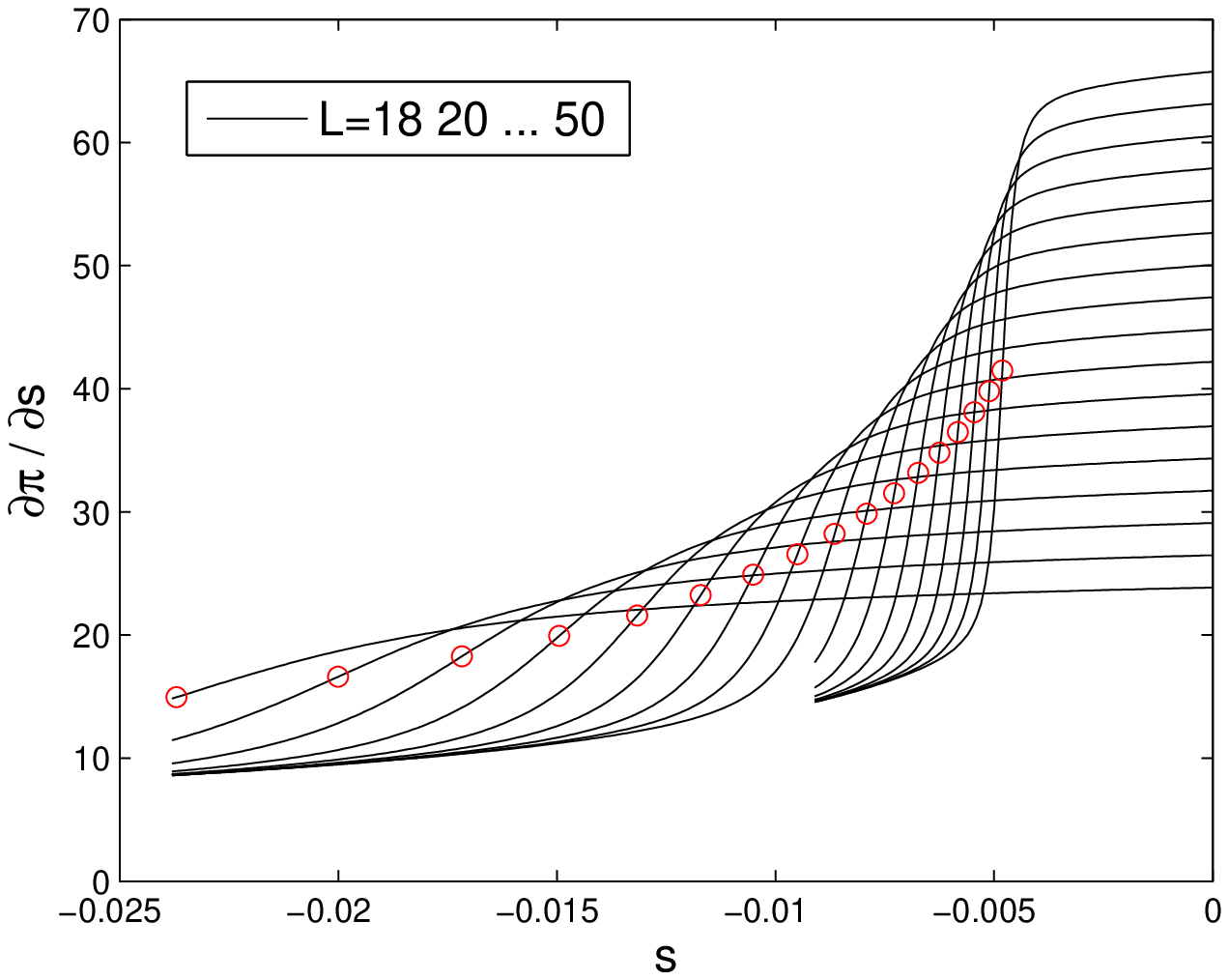}\includegraphics[width=8.0cm]{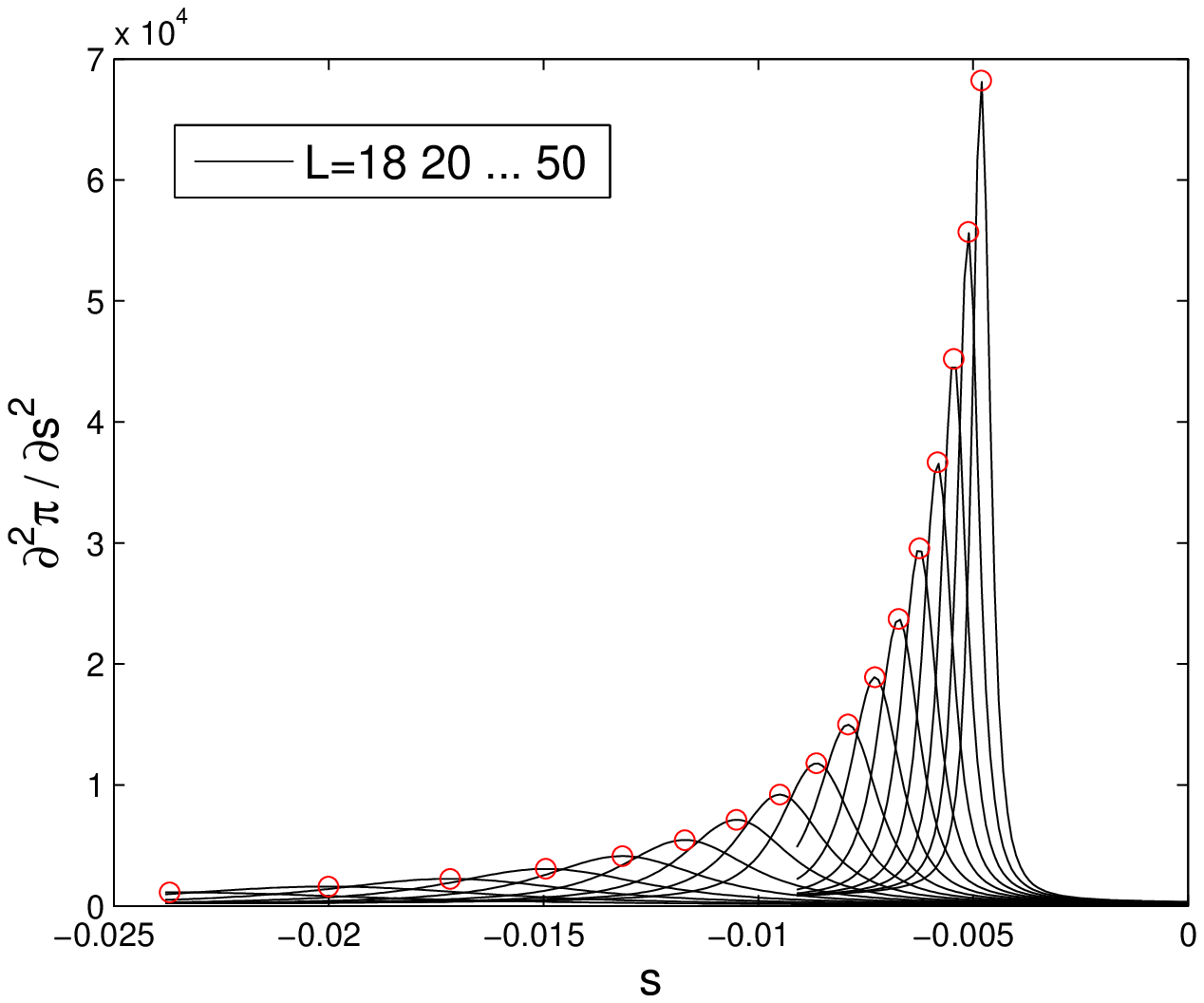}
\caption{\label{fig2} (Color online) Average $s$-weighted activity (left) and variance of the activity (right) as a function of $s$ for different $L$-values ($\lambda=4.3$). The red circles indicate the $s$-values were the variance reaches its peak.}
\end{figure}

The location of the maxima in the variance leads to finite size estimates $s_c(L)$ of the location of the transition. Our results for these are shown in Figure 3. As proven in section 2, any phase transition in the infinite system must occur at some $s_c \geq 0$. Because of the considerable curvature that is still present when plotting $s_c$ versus $1/L$ it is not possible to obtain very precise estimates of the location of $s_c$. The available data do indicate however, that for $\lambda \geq \lambda_c$, $s_c=0$ whereas for $\lambda < \lambda_c$, $s_c$ has a small positive value. For example, for $\lambda=2.3$, we estimate $s_c=0.005 \pm 0.001$. 

In the next section, we will show that the average activity is proportional to the density $\rho$ of particles.
The (ordinary) contact process, i.e. at $s=0$, always reaches the absorbing state ($\rho=0$) when $\lambda < \lambda_c$. Hence, it is not surprising that one needs a finite value of $s$ for the model to turn to the active state. On the other hand, when $\lambda>\lambda_c$, the average density is nonzero at $s=0$. Since the transition has to occur at $s\geq 0$, it must be located at $s_c=0$ exactly. 
\begin{figure}[t]
\begin{center}
\includegraphics[width=10.0cm]{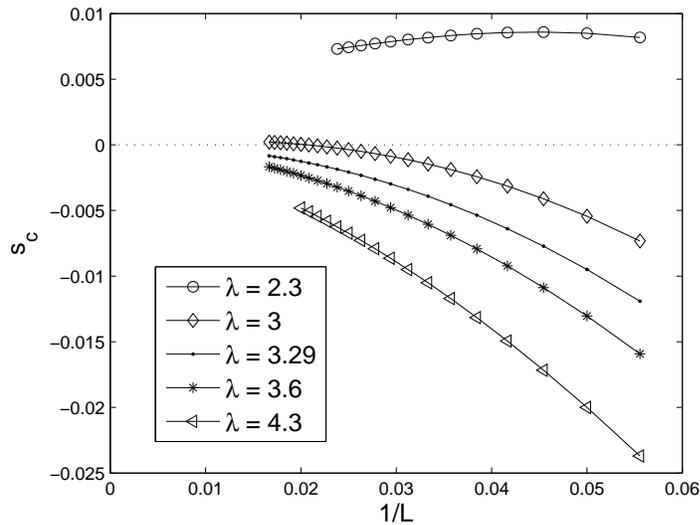}
\end{center}
\caption{\label{fig3} Finite size estimates of the location of the phase transition versus $1/L$ for different $\lambda$ values.}
\end{figure}

Other evidence for a non-zero $s_c$ below $\lambda_c$ comes from a real space renormalisation group (RG) calculation. Since the (generalised) generators of stochastic processes can be interpreted as Hamiltonians of quantum spin chains, it is natural to extend real space renormalisation schemes introduced in the study of quantum systems \cite{Drell76,Pfeuty82,Stella83} to stochastic systems. Such an extension is possible \cite{Hooyberghs00} and can give very precise values for the critical properties of the contact process \cite{Hooyberghs01}. Here, we present results of applying this scheme also to the generalised generator (\ref{6a}). Details of this calculation, and applications to another absorbing state phase transition, will be presented elsewhere \cite{Gorissen10}.
In the renormalisation we used blocks of two sites. While such a small block cannot give accurate numerical results for the location of critical points or for the value of critical exponents, it can give a good qualitative description of an RG-flow. Our results for this flow are given in Figure 4. The line $s=0$ is invariant under the renormalisation. Here we recover the results of \cite{Hooyberghs00}: for $\lambda$ below $\lambda_c$ the flow is towards $\lambda=0$, whereas in the opposite case, $\lambda$ flows to infinity. The behavior becomes more interesting in the extended parameter space $(\lambda, s)$. There  now exists a critical line (full red line). Starting from below this line, the RG flows to $\lambda=0$, whereas for parameter values above this line, the flow goes to infinity. It can be clearly seen, that the critical line is at $s=0$ for $\lambda \geq \lambda_c$, whereas it is at $s>0$ for $\lambda< \lambda_c$. This renormalisation calculation therefore supports the conclusions coming from the DMRG.
 
\begin{figure}[t]
\begin{center}
\includegraphics[width=10.0cm]{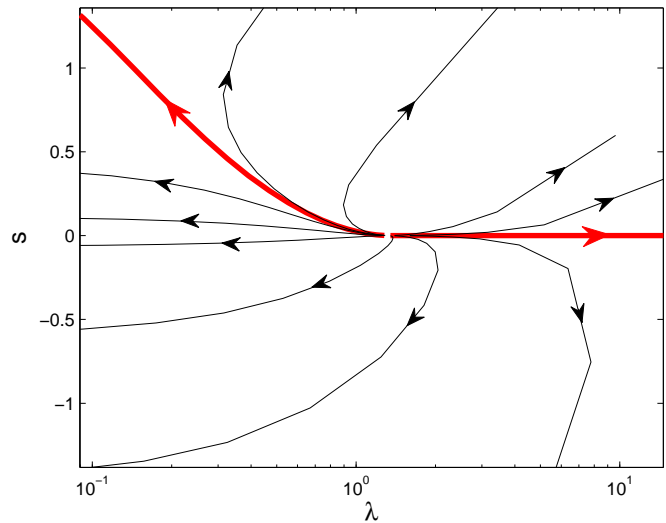}
\end{center}
\caption{\label{fig4} (Color online) Renormalisation flow for the generalised generator (\ref{6a}). The full (red) line separates regions that renormalise to $\lambda=0$ from those that renormalise to $\lambda=\infty$. The other lines show renormalisation flows for various points starting near $\lambda=\lambda_c, s=0$.}
\end{figure}

\section{Exponents and scaling behaviour}
In the previous section we have shown numerically that for every $\lambda$, the contact process undergoes a transition to an active state when $s$ exceeds a threshold $s_c$. We now investigate first the behaviour of the activity at $s=0$ as a function of $\lambda$. Then we put forward a scaling ansatz for $\pi(s,\lambda,L)$ near the critical point at $\lambda=\lambda_c, s=0$.

We begin by deriving a relation between the average activity and the density of particles. Consider a particular history which has activity $K(t)$ at time $t$. In the time interval between $t$ and $t+dt$, the activity changes to $K(t)+1$ if the model changes configuration which happens with probability
\begin{eqnarray*}
P(t) dt = \left[ \sum_{i=1}^L n_i + \frac{\lambda}{2} \sum_{i=1}^L (1-n_i)(n_{i-1}+n_{i+1})\right] dt
\end{eqnarray*}
(here again $n_0=n_{L+1}=0$ and we didn't take into account the effect of $\omega$).
The activity stays constant in this time interval if the configuration doesn't change, which happens with probability $1-P(t)dt$. Consequently,
\begin{eqnarray}
\frac{d\langle K(t) \rangle}{dt} &=& \frac{\langle K(t+dt)\rangle - \langle K(t) \rangle}{dt} \nonumber \\
&=& \left\langle \left[\sum_{i=1}^L n_i + \frac{\lambda}{2} \sum_{i=1}^L (1-n_i)(n_{i-1}+n_{i+1})\right]\right\rangle
\label{10}
\end{eqnarray}
The two-point correlation functions appearing in ({\ref{10}) can be eliminated by comparing this equation with the equation of motion for the average particle density at site $i$ which can be obtained from the master equation using standard techniques \cite{Schutz00}. One finds
\begin{eqnarray}
\frac{d \langle n_i\rangle}{dt} = \langle [H(s=0),n_i]\rangle = - \langle n_i \rangle + \frac{\lambda}{2} \langle (1-n_i)(n_{i-1}+n_{i+1})\rangle
\label{11}
\end{eqnarray}
Eliminating the two-point correlation functions between (\ref{10}) and (\ref{11}) we obtain
\begin{eqnarray}
\frac{d \langle K(t) \rangle}{dt} = 2 \sum_{i=1}^L \langle n_i \rangle + \sum_{i=1}^L \frac{d\langle n_i\rangle}{dt}
\label{12}
\end{eqnarray}
We are interested in the long time limit in which the process reaches its steady state. The first term on the right hand side then equals $2\rho L$ whereas the second term becomes zero. Hence we immediately find from the definition (\ref{3})
\begin{eqnarray}
K(\lambda,L) = 2 \rho L
\label{13}
\end{eqnarray}
The average activity is twice the average number of particles.

The particle density in the contact process has well known scaling properties near the absorbing state transition \cite{Marro99,Grassberger79}. Putting $\Delta \lambda=\lambda-\lambda_c$, one has
\begin{eqnarray}
\rho(\Delta \lambda,L) \simeq L^{-\beta/\nu_\perp} G(L^{1/\nu_\perp} \Delta \lambda )
\label{14}
\end{eqnarray}
where $G$ is a scaling function and $\beta$ and $\nu_\perp$ are known critical exponents of the contact process.
In one dimension, their values are $\beta=0.27649(4)$ and $\nu_\perp=1.09684(6)$ \cite{Jensen96}. This scaling and (\ref{13}) imply that at $\lambda_c$, the average activity behaves as a power law 
\begin{eqnarray}
K(\lambda_c,L) \sim L^\sigma
\label{15}
\end{eqnarray}
with $\sigma=1-\beta/\nu_\perp=0.7479$. We calculated finite size estimates of $\sigma(L)=\log[K(\lambda_c,L+2)/K(\lambda_c,L)]/\log((L+2)/L)$ using our DMRG technique. The results are shown in Figure 5. Using standard extrapolation technique for finite size systems \cite{Henkel88} we obtain the estimate $\sigma=0.746(2)$ in perfect agreement with the prediction made above. 
\begin{figure}[t]
\begin{center}
\includegraphics[width=10.0cm]{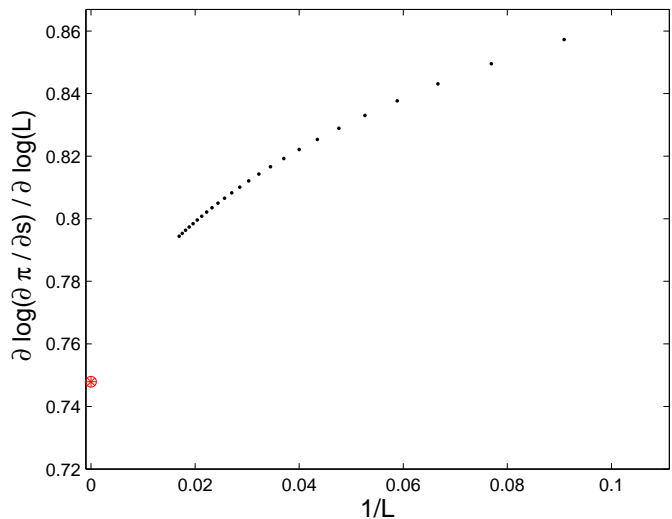}
\end{center}
\caption{\label{fig5} (Color online) Finite size estimates of the exponent $\sigma$ versus $1/L$ (for $L$ up to 60). The red circle is the prediction $\sigma=1-\beta/\nu_\perp=0.7479$.}
\end{figure}

We now turn to the variance of the activity at $s=0$. DMRG results for this quantity as a function of $\lambda$ are shown in Figure 6. The data show a clear peak whose height increases as a power of $L$. 
\begin{figure}[t]
\begin{center}
\includegraphics[width=8.0cm]{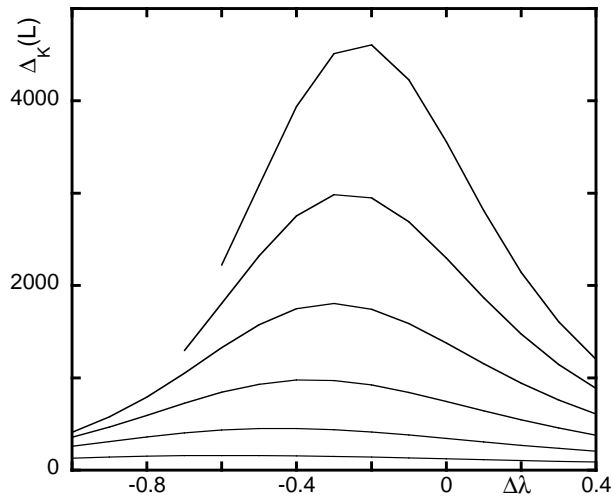}
\end{center}
\caption{\label{fig6} Variance of the activity as a function of $\Delta \lambda$ for (top to bottom) $L=44, 38, 32, 26, 20$ and $14$.}
\end{figure}
We determined the associated exponent to be $3.08(2)$. 

Given the mathematical similarity between the generating function $\pi$ and the equilibrium free energy, it is natural to put forward a scaling ansatz for $\pi$ near the absorbing state transition.
We propose therefore the following finite size scaling ansatz for $\pi$
\begin{eqnarray}
\pi(s,\Delta \lambda, L) \simeq L^{-z} F(L^{y_s}s,L^{1/\nu_\perp}\Delta \lambda)
\label{16}
\end{eqnarray}
Here $F(x,y)$ is a scaling function, and $y_s$ a new critical exponent associated with activity fluctuations. The exponent $z$ is the dynamical exponent $z=\nu_\parallel/\nu_\perp$. It replaces the dimension of space $d$ that appears in the scaling of the free energy density, because $\pi$ is a quantity {\it per unit of time}. For the one-dimensional contact process $z=1.5805$. From the ansatz (\ref{16}), we predict that the average activity at the transition scales as $L^{-z+y_s}$. Comparing with (\ref{15}) we obtain
\begin{eqnarray}
y_s=z+1-\beta/\nu_\perp=2.3284
\label{17}
\end{eqnarray}
The scaling (\ref{16}) then predicts that the variance of the activity at criticality should scale as $L^{-z+2y_s}$ where the exponent should equal $3.0763$. This value that is fully consistent with the numerical value found above. The DMRG results therefore fully support the ansatz (\ref{16}).
From this ansatz, one can derive the scaling behaviour of various quantities. For example, the quantity $K(s,L)=\partial \pi/\partial s (s,\lambda_c,L)$ should scale as $L^{-z+y_s} G(L^{y_s}s )$. Scalings like this are well satisfied (see figure 5 in \cite{Gorissen09}).

\begin{figure}[t]
\begin{center}
\includegraphics[width=10.0cm]{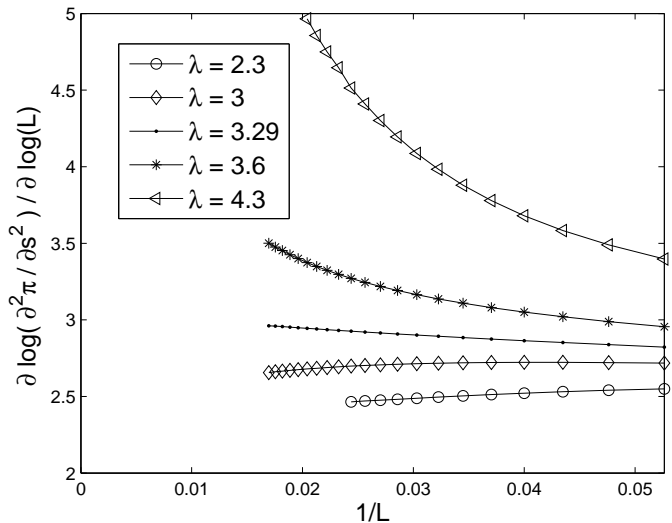}
\end{center}
\caption{\label{fig7} Logaritmic derivative of the variance (at $s_c(L)$) as a function of $1/L$ for various $\lambda$-values.}
\end{figure}
We have further investigated the behaviour of the variance of the activity along the transition line $s_c(\lambda)$. For this quantity we find a distinct behaviour below and above $\lambda_c$. 
Herefore, we go back to our data on the variance of the activity as a function of $s$ (see Figure 2). 
In Figure 7, we plot our results for $\partial \log(\partial^2 \pi/\partial s^2 (s_c(L)))/\partial \log L$ for various $L$-values. The derivative is a discrete one, as obtained from a comparison of results for system size $L$ and $L+2$. When the variance grows as a power law, this quantity should approach a constant for large $L$. This seems to be indeed the case for $\lambda \leq \lambda_c$. 
At $\lambda_c$, an extrapolation leads to an exponent value of $3.05(2)$, consistent with the value obtained above from the behavior as a function of $\lambda$ at $s=0$.
An extrapolation of the data for $\lambda=3$ gives an exponent $2.00(2)$, which would imply that the variance of the activity per site is constant. For $\lambda=2.3$, the exponent is somewhat higher ($\simeq2.4$). This could be a genuine effect, or the result of the DMRG being less stable farther away from $\lambda_c$. In contrast, in the active region (and especially for $\lambda=4.3$), it is clear that the behavior of the variance of the activity is not consistent with a power law. The data can be fitted better with an exponential growth of the variance. Such a behavior is somewhat unexpected but not impossible in principle.

\section{Conclusions}
In this paper we studied the contact process and weighted its histories according to their activity. We studied the phase diagram of the model as a function of the parameters $s$ and $\lambda$ and found that when $s$ is increased the model undergoes a phase transition between an absorbing state with zero activity and an active state. The well known absorbing state phase transition at $s=0$ is thus seen in a broader context. The transition is at $s_c>0$ for $\lambda < \lambda_c$ while it occurs at $s_c=0$ when $\lambda \geq \lambda_c$. Our results were obtained using a density matrix renormalisation group approach that allows us to calculate the full cumulant generating function of the activity. 

This generating function can also be obtained from simulations based on a cloning algorithm \cite{Giardina06,Lecomte07b}. The DMRG approach has the advantage that it gives in principle numerically exact results. The cloning algorithm is less accurate, but can easily be extended to higher dimensions, which for the DMRG is more difficult. Some preliminary results for the contact process using the cloning algorithm were reported in \cite{Lecomte07b}. These authors study a somewhat different model in which particles can be created at every site (and not just at the boundaries as in our model). Evidence for a transition at an $s_c \neq 0$ was given but no systematic study of the phase diagram and the critical exponents was presented.

In section 2, we showed that the $s$-weighted average activity (per site) is zero for $s<s_c$. Since for $\lambda>\lambda_c$, $s_c=0$ and since at $s=0$ the average activity (per site) equals $2\rho$, we conclude that in this range of $\lambda$ values the transition between non-active and active phases has to be first order.
We are unable to give definite claims on the order of the dynamical transition for $\lambda<\lambda_c$. The order of the transition can be determined from the probability distribution of the activity. Unfortunately, it is not possible to get this function unambiguously from the cumulant generating function when a first order transition is present \cite{Touchette09}. 

We have shown that the cumulant generating function $\pi$ has a scaling form close to $\lambda=\lambda_c,s=0$. The scaling form is a natural extension of that obeyed by the free energy of equilibrium systems near a critical point. A similar scaling was found to hold for the cumulant generating function of the current in the totally asymmetric exclusion process \cite{Gorissen09}. It would be of interest to investigate whether the same scaling form applies to the dynamical transition in (spin) glass models. In one dimension, kinetically constrained models do not show a transition (unless at $T=0$) but transitions in the equilibrium state of glass models can occur in higher dimensions. The validity of the scaling ansatz in these cases would enforce the link between the contact process and models for glassy dynamics.

\ack
We thank M Gorissen, E Pitard and J Tailleur for useful discussions.

\appendix
\section{Some details on the DMRG approach}
Assume one wants to know a particular eigenstate 
$|\psi\rangle$ of the Hamiltonian for a quantum spin chain, together with its associated eigenvalue. Often this is the ground state. This state of interest will be called {\it the target state}. Consider a system of size $L$ for which the target state has to be determined. For a spin $1/2$-system, this involves the diagonalisation of a matrix in a $2^L$ dimensional space. When $L$ is too big this cannot be done exactly. The idea is then to 'project' the Hamiltonian into a space of lower dimension. The basis vectors spanning this lower dimensional space are determined from a reduced density matrix of the target vector. We now briefly describe the procedure used to determine these basis vectors in the case of quantum systems since this is necessary to explain the modifications that have to be made for stochastic systems.

As a first step in the DMRG, the system is divided into two parts. One is called the system block, the other the environment. The whole system is called the superblock. Both parts are taken of equal size and the number of degrees of freedom for each is $n=2^{L/2}$. The orthonormal basis vectors representing the respective configuration space will be denoted by $|i\rangle$ and $|j\rangle$. Our aim is to reduce the number of states $n$ used to describe the system block without changing the target state $|\psi\rangle$ of the superblock. More precisely, assume one wants to reduce $n$ to $m<n$. The $n$ vectors $|i\rangle$ are to be replaced by $m$ vectors $|u^\alpha\rangle, \alpha=1,\ldots,m$. In combination with the environment states $|j\rangle$, these vectors should be able to make an accurate representation $|\tilde{\psi}\rangle$ of the target state $|\psi\rangle$ 
\begin{eqnarray}
|\psi\rangle \approx |\tilde{\psi}\rangle = \sum_{\alpha=1}^m \sum_{j=1}^n a_{\alpha,j} |u^\alpha\rangle|j\rangle
\label{A1}
\end{eqnarray}
By this we mean that one needs to minimise 
\begin{eqnarray}
S=\left| |\psi\rangle - |\tilde{\psi}\rangle \right|^2
\label{A.2}
\end{eqnarray}
with respect to all $a_{\alpha,j}$ and $|u^\alpha\rangle$. The solution of this minimisation problem is the heart of the DMRG and is given in terms of the density matrix $\rho=|\psi\rangle\langle \psi|$ of the target state of the superblock. From this matrix one constructs the reduced density matrix on the system block
\begin{eqnarray}
\tilde{\rho}_{ii'}= \sum_{j=1}^{n} \langle i|\langle j|\ \rho  \ |i'\rangle |j\rangle
\label{A.3}
\end{eqnarray}
$\tilde{\rho}$ is an Hermitian operator with $n$ eigenvalues. It can then be shown that the $m$ eigenvectors of $\tilde{\rho}$ with largest eigenvalue are the vectors $|u^\alpha\rangle$ in the solution of the minimisation problem. This result forms the basis of both the infinite size and finite size algorithms used in practical DMRG calculations \cite{White93}. 

When $H$ is the (generalised) generator of a stochastic process, left and right eigenvectors are different. In principle there are different ways to adapt the DMRG approach. We followed \cite{Carlon99} in using both the left and right eigenvector associated with the largest eigenvalue of $H(s)$ as target state. In particular, if $|\psi_0\rangle$ is the right eigenvector (and $\langle \psi_0|$ its transpose) and $\langle \varphi_0 |$ the left eigenvector (with transpose $|\varphi_0\rangle$) we used as density matrix the combination
\begin{eqnarray}
\rho=\frac{1}{2} \left[ |\psi_0\rangle\langle \psi_0| + |\varphi_0\rangle\langle\varphi_0|\right]
\label{A.4}
\end{eqnarray}
This matrix is symmetric. Its reduced density matrix $\tilde{\rho}$ is calculated as in (\ref{A.3}) and the rest of the DMRG algorithms are not modified.

\end{document}